# From the first observations of cosmic rays to the physics of relativistic nuclei

P.I. Zarubin[1,2] and A.A. Zaitsev[1,2]

[1]Joint Institute for Nuclear Research

[2]Physical Institute RAS (FIAN)

Research of cosmic rays at the Physical Institute of the USSR Academy of Sciences resulted in the construction of the JINR Synchrophasotron. For this purpose the Electrophysical Laboratory of the USSR Academy of Sciences was founded in 1953, which became part of JINR in 1956 as the High Energy Laboratory. The initial milestones to develop experiments at the Laboratory on the Synchrophasotron are presented. Leaders and key participants in the experiments are highlighted, as well as the lessons and results relevant today.

High-energy particle accelerators are sometimes figuratively called "Pyramids of the 20$^{th}$ century". These largest instruments for understanding the micro world deserve this comparison not only for their impressive size, but also for the significant impetus given to collective creativity in science and technology. Our era has a chance to be called in the human history as the "cultural layer of synchrotron builders." The longevity of installations of this class is based on the opportunity of their qualitative improvement while deepening fundamental research carried out on them and developing applied studies.

In 2023 the JINR Laboratory of High Energy Physics named after V. I. Veksler and A. M. Baldin celebrated the 70$^{th}$ anniversary of its foundation. Several generations of its employees and colleagues in the JINR Member States have traveled a glorious and dramatic path. Its milestones were the first experiments on particle physics at the Synchrophasotron, research in relativistic nuclear and polarization physics, and the construction of a specialized synchrotron based on superconductivity - the Nuclotron. Since the beginning of the 70s the experiments at the cutting edge of particle physics were performed at the Institute of High Energy Physics (USSR), Fermi National Accelerator Laboratory (USA), and the European Organization for Nuclear Research (CERN).

A long-term perspective at VBLHEP has been opened up with the NICA megaproject for studying quark-gluon degrees of freedom of nuclear matter using the extracted and colliding beams of heavy relativistic nuclei and polarized protons at the energy of several GeV per nucleon. The phased development of the project expands the opportunities of experiments in nuclear physics and radiation biology, as well as in the related applied areas. An important milestone at NIKA in 2023 was the acceleration of xenon nuclei to 3.8 GeV per nucleon. At present the accelerator cascade includes an electron beam source, a new linear heavy ion accelerator HILAC, a new superconductive booster and the Nuclotron with beam extraction systems for VBLHEP experiments. This large-scale development would have been hardly possible without the material and intellectual capital accumulated over the past decades. This capital includes the priority physical results obtained on the accelerator and experimental base and also the achievements in cosmic ray physics of the Soviet period.

The organizational "start of reference" of VBLHEP is taken to be the formation in 1953 of the Electro physical Laboratory of the USSR Academy of Sciences (EFLAN) as an institute within the Department of Physical and Mathematical Sciences to perform research in the field of high-energy physics at the Synchrophasotron under construction. EFLAN was headed by V.I. Veksler - a recognized leader in cosmic ray research in the 30s and 40s, as well as the leader of the construction of the first synchrotrons based on the phase stability principle he discovered. In 1956, EFLAN joined JINR as a laboratory with the name Laboratory of High Energy (LHE). EFLAN replaced the TDS-533 - the technical directorate of construction, where future VBLHEP employees were accepted. When VBLHEP merged with the Laboratory of Particle Physics in 2008, the full name of VBLHEP was returned.

One of the key moments in the emergence of high-energy physics was the observation in 1927 by D.V. Skobeltsyn of tracks of particles with momenta above 20 MeV/$c$ in a cloud chamber placed in a magnet, which indicated the corpuscular nature of cosmic radiation. The study of cosmic rays continued under the leadership of D.V. Skobeltsyn at the Atomic Nucleus Laboratory at the Physical Institute of the USSR Academy of Sciences [1]. New detection methods were required to solve fundamental problems. To study the nature of cosmic rays, V.I. Veksler was involved in the complex expedition of the USSR Academy of Sciences to Elbrus. Veksler became a recognized expert on Geiger counters and in 1937 N.A. Dobrotin [2] noted the following:

"V.I, Veksler reported to us on the application of the original technique he developed to the study of the cosmic rays. It consists of using gas proportional amplifiers operating according to a coincidence circuit to count particles. This makes it possible to determine not only the number of particles that have passed through these counters, but also to measure the ionization they produce. V.I, Veksler worked with such a setup this summer during an expedition to Elbrus. It turned out that at an altitude of 4200 m above the sea level there are easily absorbed and highly ionized particles. At the sea level, the number of these particles is noticeably smaller than at the height of Elbrus. Their number is so small that these observations cannot be reconciled with the assumption of the presence of an intense proton component in cosmic rays. In addition, V.I, Veksler obtained indications of the secondary nature of these particles. A number of authors have pointed out the existence of these particles before. But it is for the first time they have been discovered with such clarity. Thus, already the first experiments with proportional gas amplifiers gave very valuable results. And there is no doubt that the further use of this technique will make it possible to achieve very significant success, both in the study of heavy particles and showers."

Research on cosmic rays was continued in the FIAN expedition to the Pamirs, which began on behalf of the Soviet leadership in the spring of 1944. "The discovery of electron-nuclear showers during the Pamir expedition and the identification of the nuclear cascade process allowed one to reveal the mechanism that represents the initial link in the chain of cosmic rays developing in the atmosphere. It was shown that electromagnetic cascade processes, with which the essence of widespread air showers had previously been associated, are only an external manifestation of nuclear interactions caused by high-energy particles such as mesons, protons, and heavier atoms." [3].

Future LHE professors K.D. Tolstov, A.L. Lyubimov, M.I. Podgoretsky went through Veksler's school on the Pamir expedition, who headed the first scientific teams at the Synchrophasotron. The tasks of post-war reconstruction and defense of the country did not encourage academic expenditures in the harsh environment of the 40s. However, due to the fact that the non-military nature of this area was not obvious at the initial stage, the fundamental research in this area received unprecedented support within the framework of the USSR Atomic Project. The very logic of the research required laboratory conditions for a targeted study of the particle production, identifying their internal structure, and searching for unknown particles and antiparticles. The landmark phase stability principle (auto-phasing) [4], proposed independently by V.I. Veksler and E. McMillan in USA, solved the problem of constructing accelerators in the cosmic energy scale.

At the end of the 40s under the leadership of V.I. Veksler, electron synchrotrons were made: at 30 MeV for studying photonuclear reactions (S-3, "troika") and at 250 MeV (S-25) [5, 6]. The dissertation of I.V. Chuvilo, dedicated to the production of neutrons by γ-quanta was the first one on S-3. At S-25, by means of the Ilford nuclear emulsion, the formation of π-mesons by γ-quanta was discovered (Fig. 1) [6]. This marked the beginning of the physics of electromagnetic interactions of hadrons, figuratively called then "nuclear properties of light." The main directions were: the study of the production of charged and neutral mesons on nucleons, the polarizability (deformability) of nucleons in Compton scattering of γ quanta. In general, the experiments demonstrated the opportunity of describing hadronic physics based on quantum field theory.

Young physicists who later took leading positions in scientific centers began working at the S-25: M.I. Adamovich, A.M. Baldin, A.S. Belousov, B.B. Govorkov, V.I. Goldansky, A.N. Gorbunov, Yu.M. Ado, S.P. Denisov, R.M. Lebedev, M.F. Likhachev, V.I. Moroz, A.P. Onuchin, L.N. Strunov, E.I. Tamm, I.V. Chuvilo, L.N. Starkov. This list is almost certainly not complete. Mastering the injection and acceleration of slow protons required to develop a model MKM synchrotron (directed by V.A. Petukhov and L.P. Zinoviev), on which protons were accelerated to 160 MeV (Fig. 2). Emphasizing this success, V.I. Veksler said: "… the Synchrophasotron "at the Volga" will definitely work." Later, the MKM was converted into the S-60 electron accelerator and served at the Lebedev Physical Institute as a source of synchrotron radiation until dismantled in the 2000s.

At the Lawrence Berkeley Laboratory (USA), under the leadership of E. McMillan, a weak focusing synchrotron was constructed to accelerate protons to 6 GeV. Thanks to the developments carried out at the laboratories under the leadership of V.I. Veksler and A.L. Mintz, the Soviet response in the spring of 1957 was the commissioning of the Synchrophasotron with the proton energy of 10 GeV (Figs. 3 and 4). The physical justification of the technical project was developed in 1950 by the group of V.I. Veksler which included A.M. Baldin, A.A. Kolomensky, A.P. Komar, V.V. Mikhailov, and M.S. Rabinovich. It is worth noting that, based on their research experience, the "fianovites" led by M.A. Markov, insisted on the electronic version, which took place in the 60s under the leadership of E.I. Tamm, as S-25R in the Department of High Energy Physics of the Lebedev Physical Institute in Troitsk. In the case of the Synchrophasotron, the point of view of physicists associated with I.V. Kurchatov's Laboratory No. 2 who advocated the proton version. This choice determined the scientific fate of LHE 20 years later: the emergence of relativistic nuclear physics on the basis of the Synchrophasotron.

This way began the era of peaceful competition in the physics of the microworld. Soon it became possible to build science cities and accelerator centers in the USSR, the USA and Western Europe, already open to the widest international cooperation. They can be considered general milestones of human progress over the past decades. In the autumn of the same year, the first artificial satellite was commissioned - "piccola luna russa" according to one of the newspaper headlines abroad as A.M. Baldin recalled.

In Moscow, the technological and analytical foundations of the nuclear emulsion method for relativistic particles were mastered to perform fundamental observations of relativistic particles in high-altitude exposures. Figure 5 shows a projective photograph of the interaction of a nucleus with gigantic energy in a nuclear emulsion, which resulted in a tremendous multiplicity of secondary particles. Traces in the stream of fragments can be completely resolved. The completeness and reliability of observing the traces of nuclear reactions, the unique resolution and flexibility of application of this method have remained unsurpassed till today.

The development of a domestic thick-layer nuclear emulsion before launching the Synchrophasotron became the starting point for obtaining the first physical results. It is worth recalling that this event raised doubts abroad. Thus, Nobel laureate S. Powell, who discovered the π-meson, linked his agreement to the FIAN physicists to republish his fundamental work [7] in the USSR with the provision of photographic plates that objectively confirmed the success of the accelerator. Photographs of characteristic interactions were included in the Russian publication (Fig. 6) [7]. The analysis of the first exposures made it possible to provide an overview of the topology of nuclear stars and carry out angular measurements of the emerging traces [8-10]. Till now, the observations of the complete destruction of target nuclei and the coherent production of mesons in events not accompanied by tracks of target fragmentation have remained unique. Exposures of emulsion stacks continued in 60es with respect to elastic scattering as a source of information about the structure of protons.

The discovery of strange particle decays in nuclear emulsions irradiated in the stratosphere led to the construction of bubble chambers that covered many orders of magnitude larger detection volumes in the magnetic field. The Nobel lecture by L. Alvarez [11] allows one to get acquainted with the level of research and the scope of innovations in the United States in this direction. The solution for LHE was to make a 40 cm bubble chamber based on liquefied propane. To expose it, an external beam of 8.3 GeV/$c$ π$^-$-mesons was formed.

The analysis of 40 thousand photographic frames taken with this camera turned out to be extremely productive from the camera (Fig. 7-9). The flagship result of LVE was identification in 1960 of the decay of a previously unobserved anti-Σ$^-$-hyperon into a π+ meson and an antineutron [12, 13]. The latter manifested itself in an annihilation star (Fig. 9). At the top of the formation of the anti-Σ$^-$-hyperon there are traces of K$^-$, and decays of a pair of K$^0$-anti-K$^0$ mesons

(So much "strangeness" at once!). Such a clearly interpretable photograph was used as a symbol of particle physics in JINR publications. It is gratifying to note that this identification channel has been used recently in the ALICE experiment at the Large Hadron Collider.

A fundamental success factor was the involvement of V.I. Veksler in the Synchrophasotron as a project of national importance for two groups of graduates of the Faculty of Physics of Moscow State University. Answering skeptics, Veksler insisted: "The youth will not let me down!" One group is shown in its entirety in the photograph (Fig. 10). Among the co-authors of the first publications one meets the names of young scientists: E.N. Kladnitskaya, L.N. Strunov, E.N. Tsyganov, V.A. Belyakov, V.A. Nikitin, and A.V. Nikitin. The global level to which JINR has reached played an important part in the choice in 1964 of Dubna to host the XII International Conference (Rochester) on High Energy Physics.

To develop particle physics up to antiprotons, in 1965 the concept of the Synchrophasotron was proposed as a source of secondary particles generated on its internal targets [14]. First of all, the increase of the intensity of circulating protons by three to four orders of magnitude was required, which needed the construction of a new linear accelerator for injection. A project appeared to make a powerful concrete protection over the ring. Then it was necessary to build channels for separated beams of secondary particles, including electrostatic separators of $K$-mesons and antiprotons, which were located in the attached building 1B. A neutron channel through Building 1B to the hydrogen chamber was implemented in the fragmentation of relativistic deuterons.

The proposed paradigm turned out to be a powerful stimulus to develop experimental methods at LHE, which made experiments in relativistic nuclear physics possible in the 70s. The focus of efforts was the construction of a 2-meter propane, xenon and 1-meter hydrogen and streamer chamber. Methods of electronic experiments were developed - Cherenkov gas counters, organic plastic scintillators, multiwire spark and proportional chambers, Cherenkov total absorption detectors, nanosecond electronics, interface with computers. It required cryogenic targets, nuclear electronics, and interface with electronic computers, which were just beginning to appear. Surely this list is incomplete, and it is worth looking through the JINR annual reports of that period.

Along this path there was not only success, but also disappointments. First of all, it was not possible to obtain the expected intensity of the separated beams. At the same time, the resources and experience of LHE scientists and specialists were required for prestigious experiments at the Institute of High Energy Physics, where in 1967 the strong focusing synchrotron U-70 with the record proton energy was put into operation on the occasion of the 50[th] anniversary of the Great October Socialist Revolution. Having appeared on the new frontier of high-energy physics, this accelerator attracted the attention of not only physicists of the Soviet Union and JINR member countries, but also of CERN and the USA.

The difficulties were aggravated by the unexpected death in 1966 of V.I. Veksler, a charismatic personality who knew how to take responsibility and solve not only scientific and technical, but also organizational challenges of unprecedented complexity. I.V. Chuvilo ("Veksler's Right Hand") headed the Institute of Theoretical and Experimental Physics, where the strong-focusing synchrotron U-10 had already been operating, which served as a prototype of the Serpukhov U-70 accelerator. Proposals arose to orient LHE for the work at IHEP with the corresponding decisions regarding the Synchrophasotron and major part of the LHE staff associated with it.

To bring the situation out of the crisis and preserve the expensive scientific capital in action, it was necessary to quickly update the program and agenda of internal experiments at LHE. The conditions were determined, on the one hand, by scientific relevance and prospects, and on the other hand, - by the real reserves of the LHE and budgetary compromise with the emerging external obligations. M. A. Markov, who replaced V. I. Veksler as academician-secretary of the Department of Nuclear Physics of the USSR Academy of Sciences, recommended his student A. M. Baldin to the position of director of LHE (Fig. 11).

Having begun his work at the Lebedev Physical Institute on the physical motivation of the Synchrophasotron, A. M. Baldin by that time had received recognition as a theorist in the physics of electromagnetic interactions of hadrons and nuclei. He collaborated with LHE in the search for electromagnetic decays of vector mesons, culminating in the first observation of $\varphi \to e^+e^-$ [15, 16]. There is no coincidence that this direction became the topic of the first International seminar on problems of high energy physics, held in September 1969 under the chairmanship of A.M. Baldin [17]. In a while the topics of this forum expanded to include multiple processes, extreme nuclear fragmentation, and quantum chromodynamics. The seminar acquired the status of a major conference with significant international and Russian participation.

The starting points for A. M. Baldin in the scientific direction formulated by him and called relativistic nuclear physics, were the data that had just appeared at IHEP, demonstrating the scale-invariant behavior of the inclusive spectra of produced hadrons. On the other hand, research of deep inelastic electron scattering at Stanford (USA) pointed to a point-like (or parton) structure of nucleons. This picture was also supported by the validity of the Matveev-Muradyan-Tavkhelidze quark counting rules for elastic hadron scattering with large momentum transfers [18].

A. M. Baldin proposed to extend the hypothesis of scale invariance to the collision of relativistic nuclei, as a consequence of the point mechanism of interaction of nuclear nucleons. Testing the hypothesis required studying the production of pions in the region of limiting fragmentation, which at the energy of about 3 GeV per nucleon required going beyond the limits of nucleon-nucleon kinematics [19]. A.M. Baldin recalled a brief remark exchange with R. Feynman during the American Institute of Physics conference in 1971 that went something like this: "I know when the Feynman variable can be greater than one." Answer: "No, this is impossible... (a pause) ..., well, of course, in the nuclei!"

In the 70-80s exhaustive measurements of inclusive spectra of cumulative hadrons, including antiprotons, and baryon fragments were carried out by the group of V.S. Stavinsky at LHE and G.A. Leksin at the Institute of Theoretical and Experimental Physics in Moscow. They constituted a unique layer of the physical picture of parton degrees of freedom in nuclear matter (reviews [20-23]) and provided the basis to develop A.M. Baldin's ideas under his leadership on the self-similar behavior of multiple particle production in collisions of hadrons and nuclei based on the data from bubble chambers [24,25].

The acceleration of deuterons at the Synchrophasotron, which concluded with the discovery of the cumulative effect, became a prerequisite to start new experiments under the leadership of L.N. Strunov and L.S. Azhgirey for studying the deuteron wave function at small inter nucleon distances. The world level in the study of spin aspects in the deuteron structure was achieved due to the source of polarized deuterons made in the 80s under the leadership of Yu.K. Pilipenko. On its basis, in the 90s, experiments were started with a polarized proton target delivered from DAPhNIA (Saclay). The extracted beams of light nuclei obtained from the electron beam and laser sources made it possible to carry out survey studies using the nuclear emulsion method (Fig. 12), in a 2-meter propane chamber (Fig. 13), 1-meter bubble and streamer chambers. All the topics covered are very important and even more - deserve their own retrospective reviews.

In practical terms, experiments in relativistic nuclear physics performed in the 70s gave a vital impetus for the modernization and regular operation of the Synchrophasotron. Under the leadership of L.P. Zinoviev, a new linear injector was put into operation, and under the leadership of I.B. Issinsky, a slow extraction of the beam with the intensity 4 orders of magnitude greater than that of the internal beam in the 60s, was carried out. Thus, shielding over the accelerator has lost its relevance. Minister of Medium Engineering of the USSR E.P. Slavsky supported the initiative by A.M. Baldin to construct a new building 205 for the extracted beams, which was carried out under the leadership of L.G. Makarov (later a "locomotive" of the Nuclotron project). Having a 20-year resource, according to D.V. Skobeltsyn, the huge vacuum volume of the Synchrophasotron chamber was maintained in the working condition until the end of its operation in the late 90s.

Thus, the targeted improvement of the accelerator, the active experiments on it, and the formulation of conclusions and generalizations mutually pushed each other, attracting an ever wider circle of users. It seems that in this close relationship lies the secret of the foundation of relativistic nuclear physics and its entry to the world level and the formation of prerequisites to construct the Nuclotron in the 80s. These scientific and historical lessons by V.I. Veksler and A.M. Baldin, their predecessors, associates and followers have still remained relevant. Participation in experiments in relativistic nuclear physics at the leading accelerator centers in the world was developed on this platform. A retrospective look at the history of VBLHEP is constructive up to now, as it allows us to apply and extrapolate a kind of arrow of time. Facsimiles of almost forgotten articles and memoirs, accumulated on the website of the BECQUEREL emulsion experiment, following the publications of JINR and the Internet, allow one to feel the inspiring atmosphere of the first steps and follow the scientific growth of VBLHEP in facts.

Since VBLHEP is in the focus of the essay, the author has allowed himself not to dwell on foreign milestones, counting from the discovery of relativistic nuclei in the stratosphere in 1949 [26]. Among them is the deployment in the 70s research with relativistic heavy ions at Bevalac at the Lawrence Berkeley Laboratory (USA) [27]. Another major milestone is the physics of relativistic few-nucleon systems, which formed the agenda first on the weak-focusing SATURN (CEA, Saclay, France), continuing on the strong focusing SATURN-2 (DAPhNIA-CEA Laboratory). A retrospective review of the progress of these laboratories would be instructive. Their achievements are presented and noted in the proceedings of International

Seminars, which turned out to become a recognized international conference in the 70-80s (Fig. 14).

Our site [28] focused mainly on the ongoing experiment BECQUEREL can serve as a starting point to penetrate into the Russian history of high energy physics. Historical materials have been accumulated in the background to prevent losses or oblivion. Otherwise, there is a risk of forgetting about the impressive development of microworld physics, more than our contemporaries often know. Since direct citation here is difficult, the author is ready to help in finding both cited materials and many others, and is also interested in expanding it. For obvious historical reasons, most of the rarities on the BECQUEREL website are in Russian, and previous publications are added as they are found. A collection of abstracts from the 1964 Dubna conference has recently been added to the "papers/books" section. Through the same section, chapters of annual reports of LHE JINR since 1965 are available. A good source for independent search is the INSPIRE database, which contains materials from International conferences on high energy physics since the 50s.

The author is grateful to Prof. A.I. Malakhov for the invitation to present reports on the occasion of the 65$^{th}$ anniversary of the Synchrophasotron in 2022 and opening of the XXV International Seminar on High Energy Physics ("Baldin Autumn - 2023"). Personal work of I.G. Zarubina on the BECQUEREL site began by her initiative in 2004, has been continued to the present, and deserves sincere gratitude.

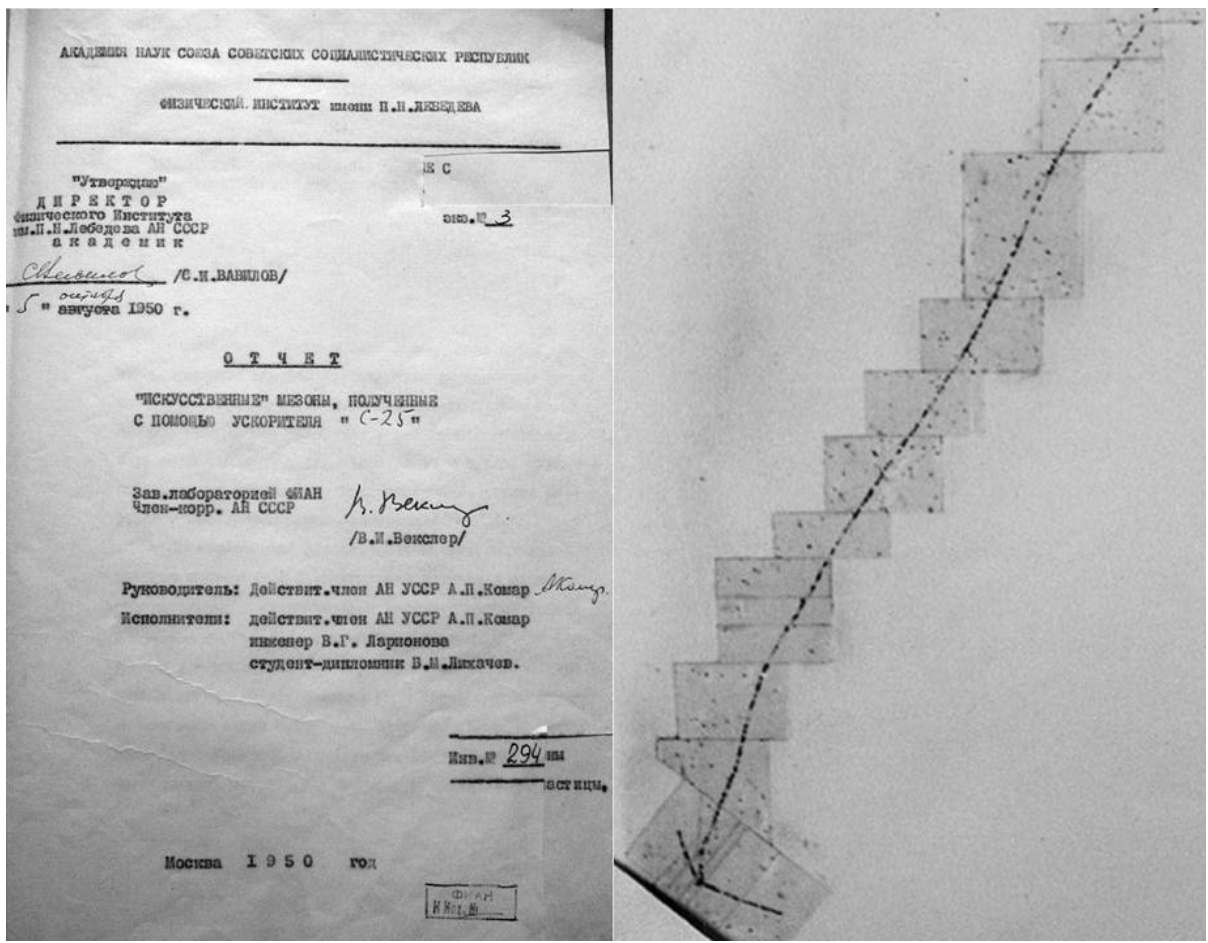

Fig.1. The title page of the Lebedev Physical Institute report and photograph from it of the tortuous trace of a π⁻-meson captured by emulsion nucleus with formation of two fragments (courtesy by S.P. Kharlamov).

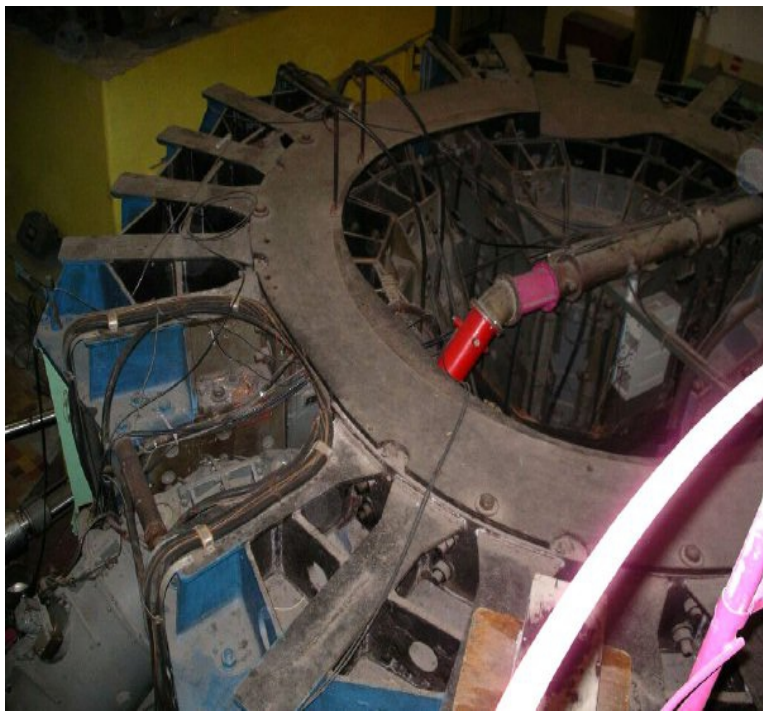

Fig.2. Photograph of synchrotron MKM (FIAN, 2007)

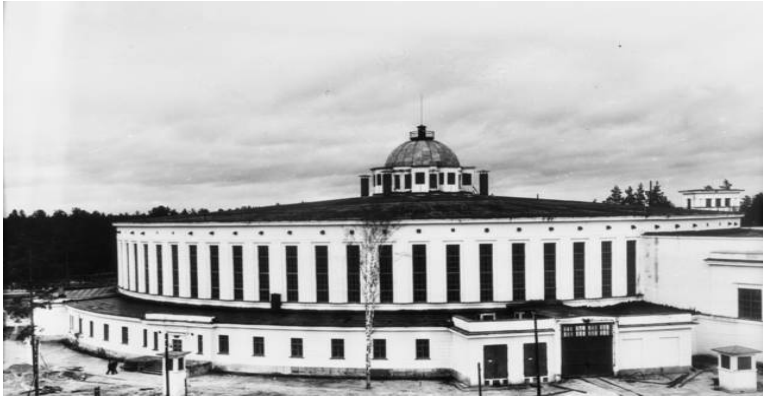

Fig. 3.The Synchrophasotron building in 1956

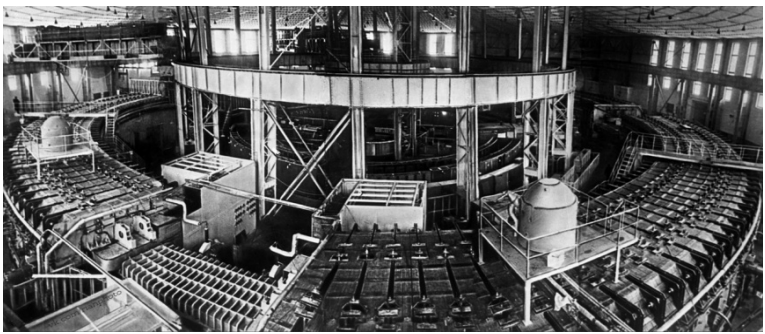

Fig.4.Inside the Synchrophasotron building (1957)

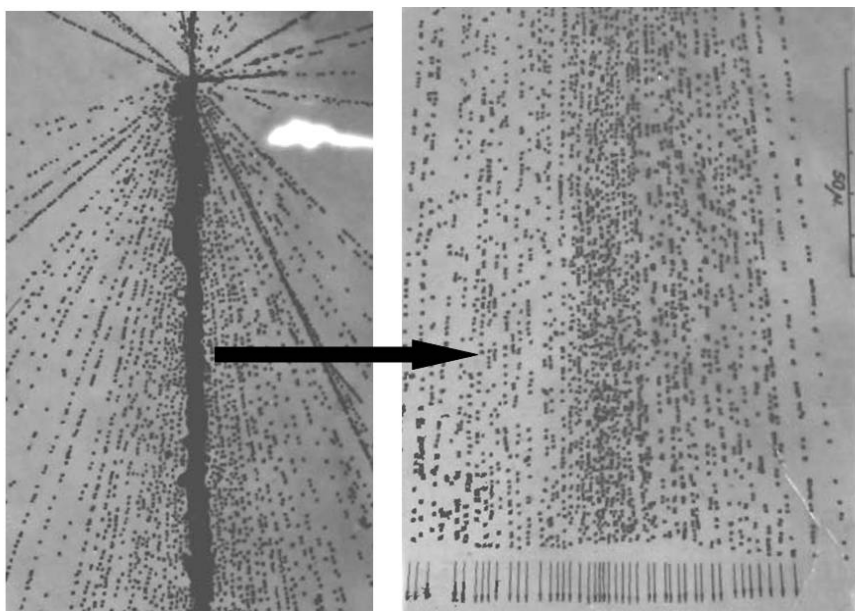

Fig.5. Macro photograph of the nucleus interaction of cosmic origin in nuclear emulsion (FIAN, early 50s, courtesy by G.I. Orlova)

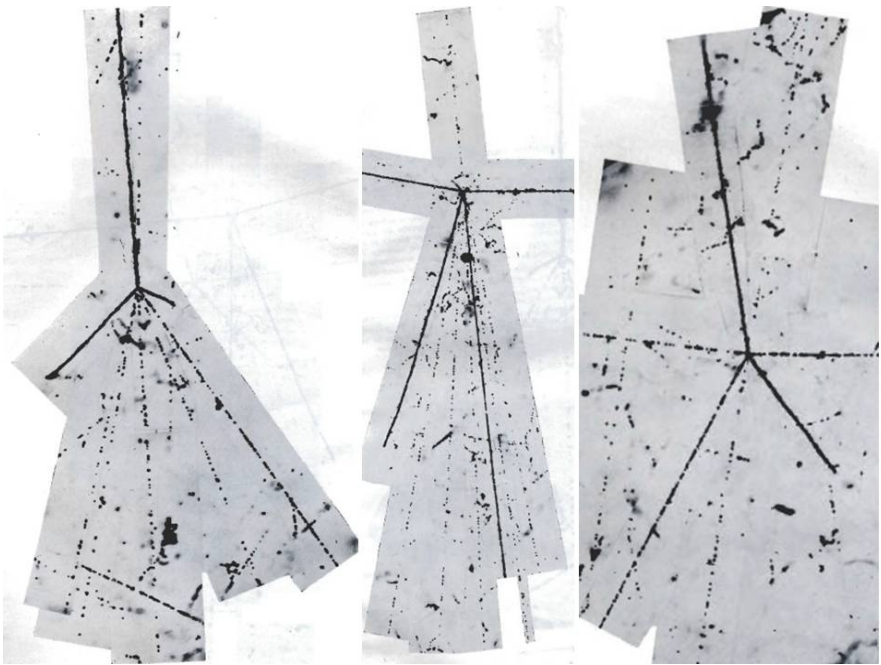

Fig.6. Photographs of interactions of the protons accelerated in the Synchrophasotron from the Russian edition [7]

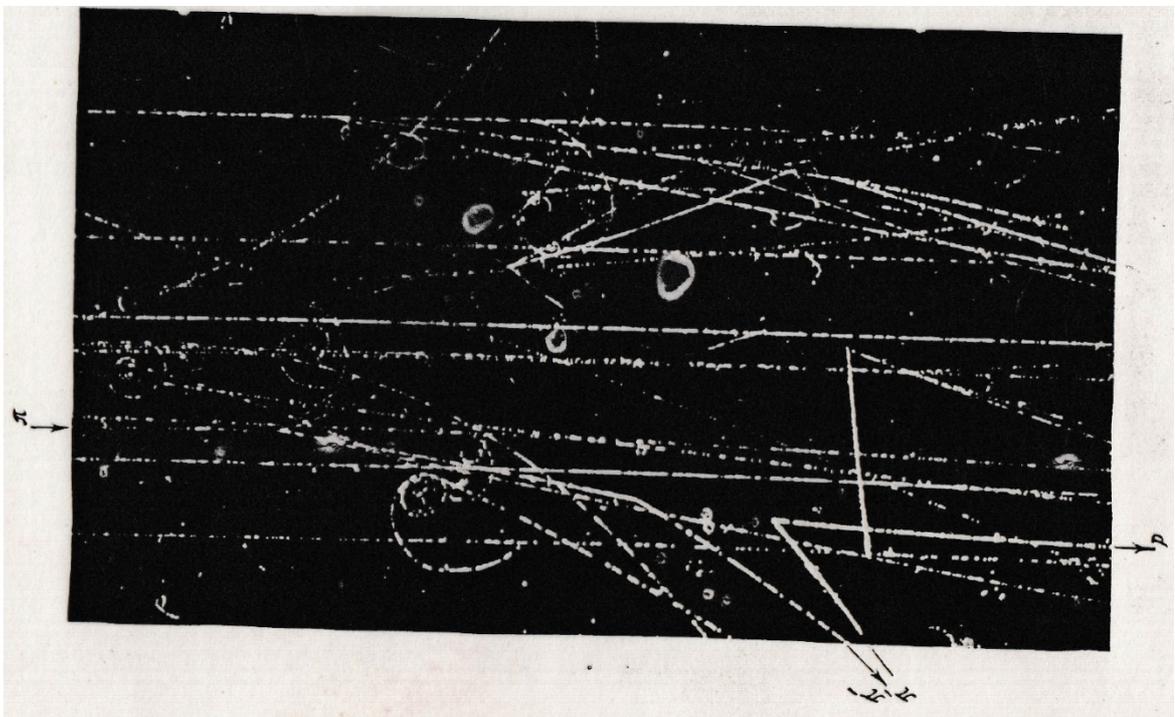

Fig.7. Photograph in the bubble chamber of production of Ξ- hyperon decaying into a Λπ- pair (courtesy of A.A. Kuznetsov)

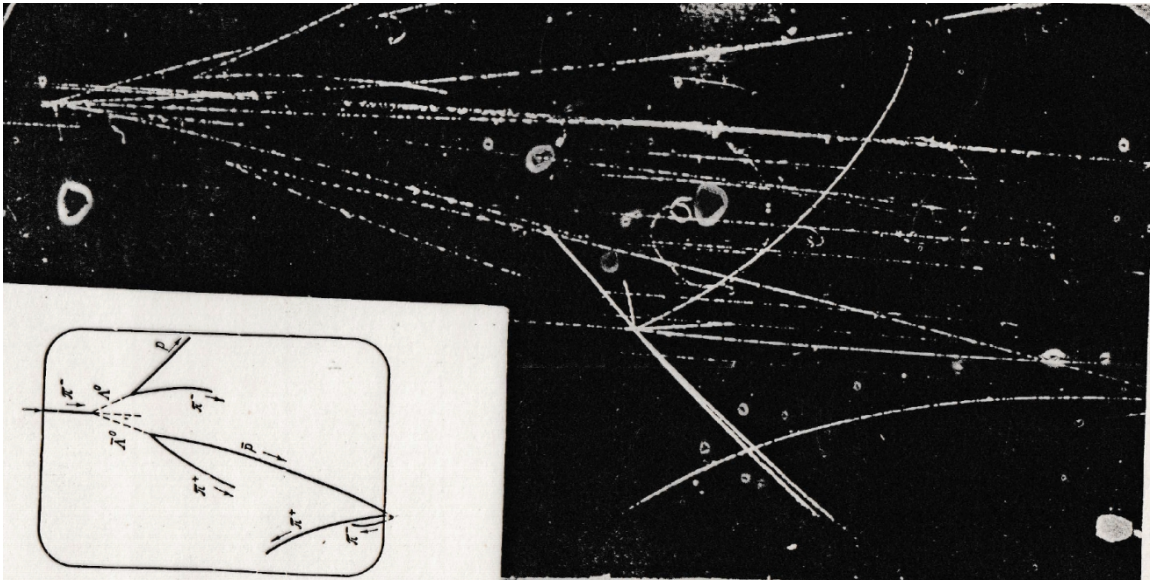

Fig. 8. Photograph in the bubble chamber of the rare event of formation and decay of Λ and anti-Λ hyperon pair (courtesy of A.A. Kuznetsov).

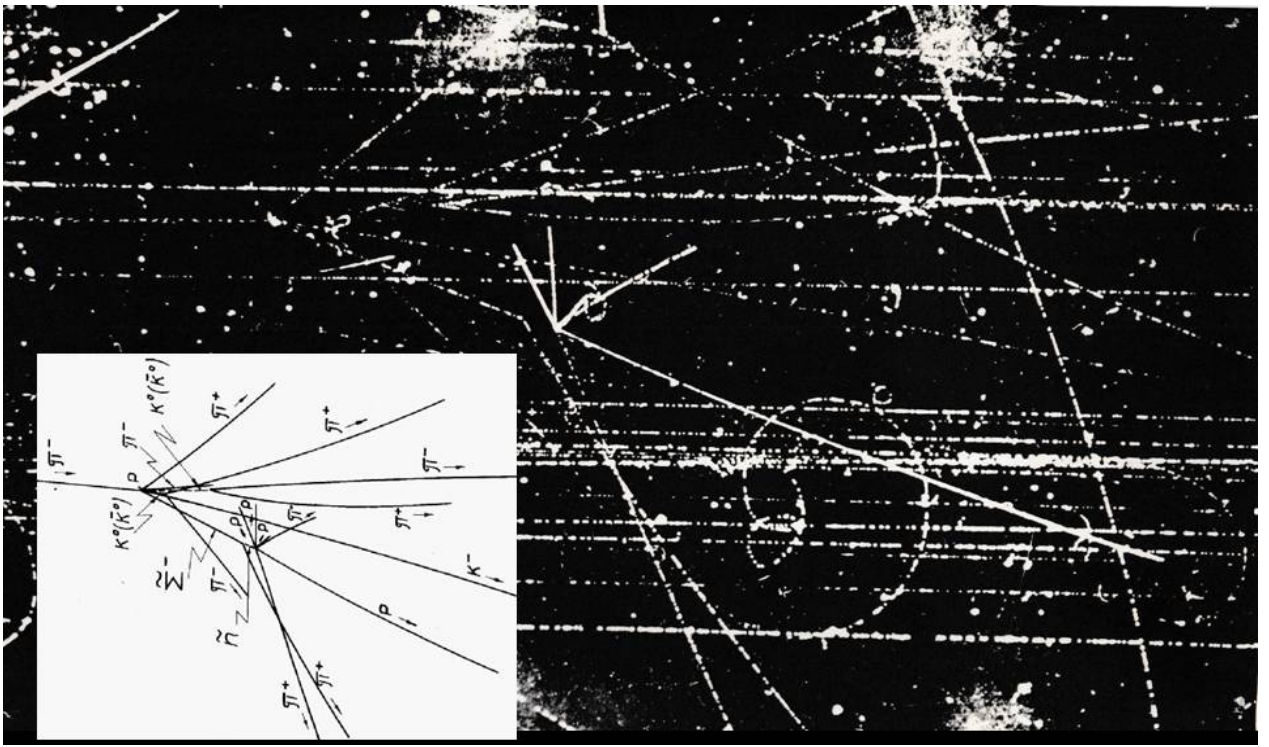

Fig. 9. Photograph in the bubble chamber of formation and decay of anti-Σ- hyperon into antineutron manifested in an annihilation star and a π+ meson (courtesy by A.A. Kuznetsov).

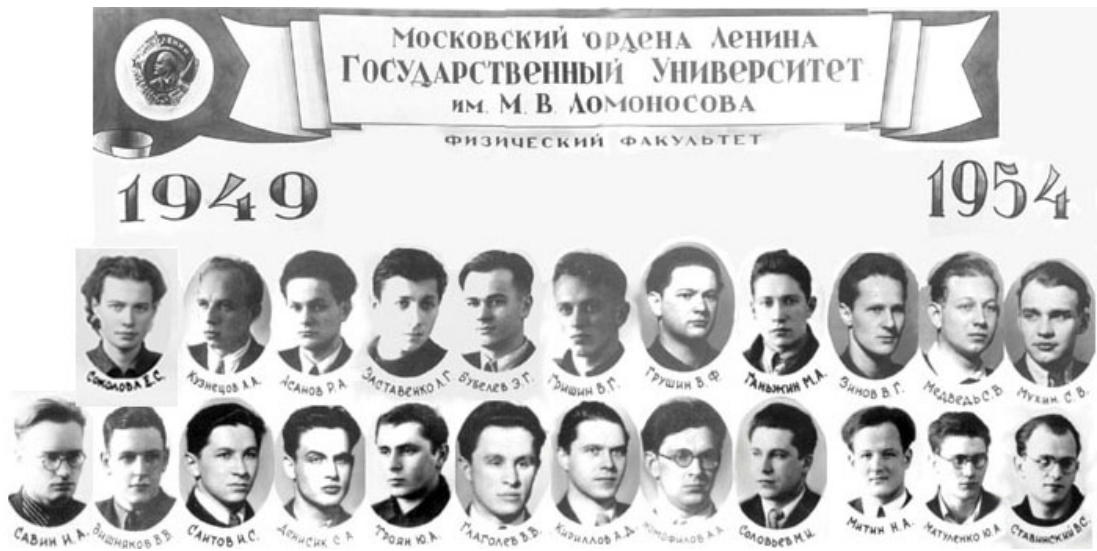

Fig. 10.The graduates of one of two groups of the Faculty of Physics of Moscow State University invited by V.I. Veksler: E.S. Sokolova (Kuznetsova), A.A. Kuznetsov, R.A. Asanov, L.G. Zastavenko, E.G. Bubelev, V.G. Grishin, A.F. Grushin, M.A. Ganzhin, V.G. Zinov, S.V. Medved, S.V. Mukhin, I.A. Savin, V.V. Vishnyakov, I.S. Saitov, S.A. Denisik, Yu.A. Troyan, V.V. Glagolev, A.D. Kirillov, A.A. Nomofilov, M.I. Soloviev, N.A. Mitin, Yu.A. Matulenko, V.S. Stavinsky (courtesy by V.V. Glagolev).

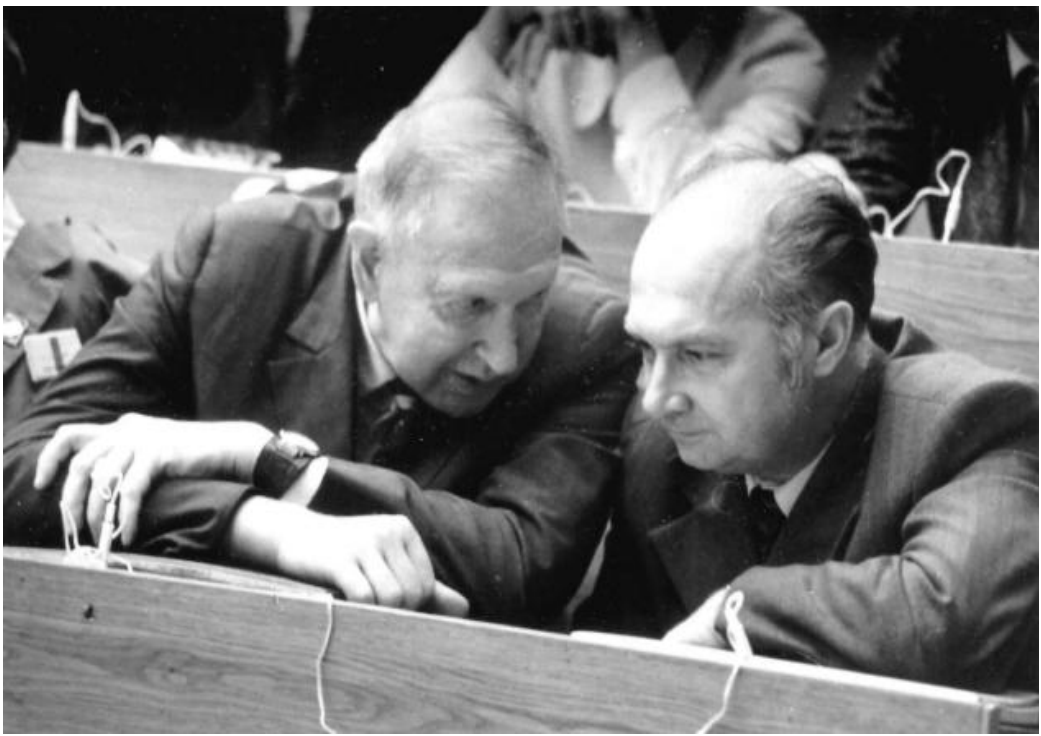

Fig.11. M.A. Markov and A.M. Baldin (photo by Yu.A. Tumanov).

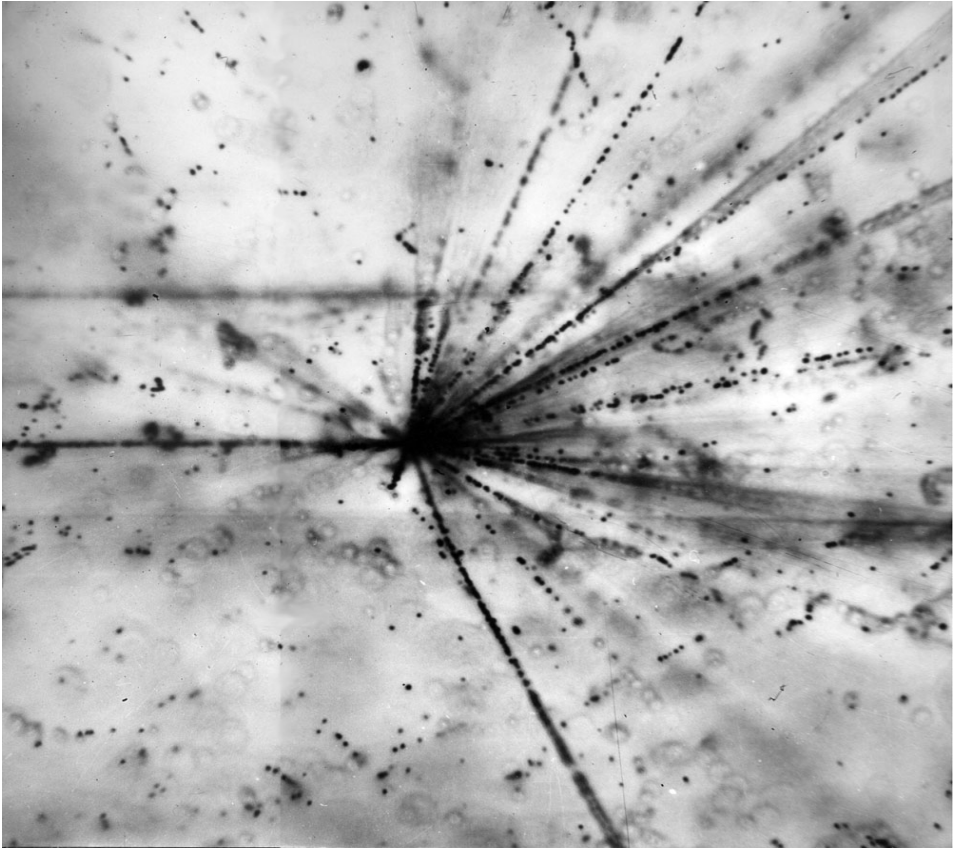

Fig.12. Photograph of the central collision of $^{12}$C nucleus of 3.65 GeV per nucleon in nuclear emulsion (courtesy by K.D. Tolstov).

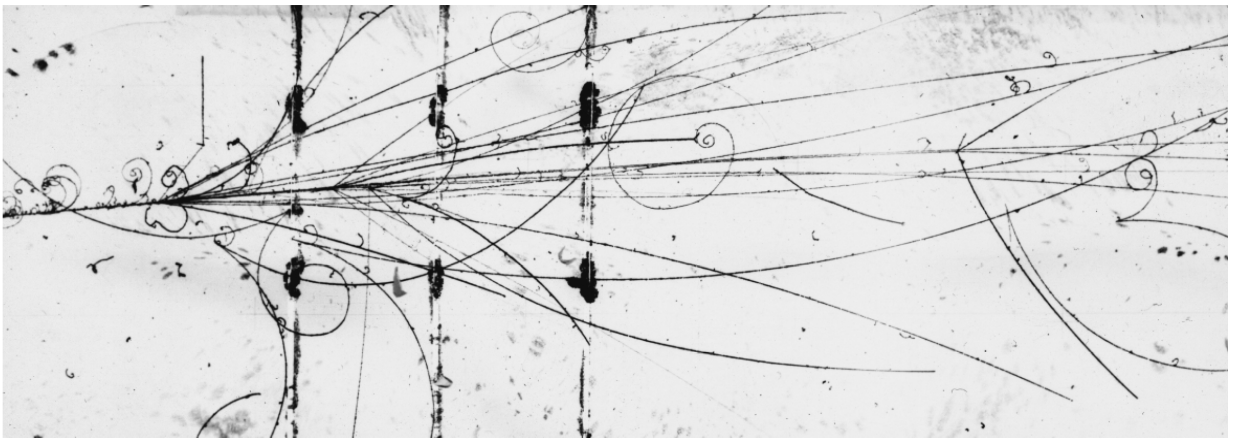

Fig.13. Photograph of central collision of C nucleus of 3.65 GeV per nucleon in the propane bubble chamber (courtesy by M.I. Soloviev).

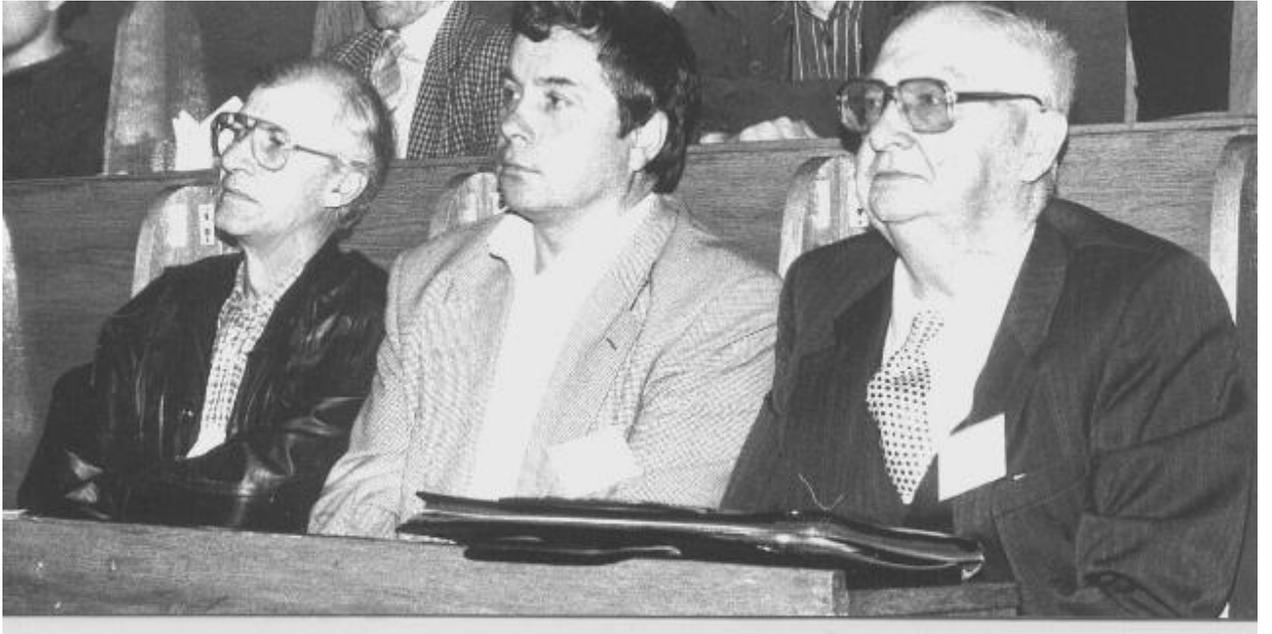

Fig.14. A.D. Kovalenko, A.I. Malakhov, and A.M. Baldin at the International Seminar on Problems of High Energy Physics (90s, photo by Yu. A. Tumanov)